\documentclass[aps,twocolumn,groupedaddress,floatfix]{revtex4}

\usepackage{epsfig}

\begin{document}

\title{Path-integral Monte-Carlo simulations for electronic dynamics
on molecular chains:\\
I. Sequential hopping and super exchange}

\author{Lothar M{\"u}hlbacher}
\email[email:~]{lothar.muehlbacher@physik.uni-freiburg.de}
\author{Joachim Ankerhold}
\author{Charlotte Escher}
\affiliation{Physikalisches Institut,
 Albert-Ludwigs-Universit{\"a}t,
 D-79104 Freiburg, Germany}

\date{Date: \today}

\begin{abstract}
An improved real-time quantum Monte Carlo procedure is presented and applied to
describe the electronic transfer dynamics along molecular chains. The model
consists of discrete electronic sites coupled to a thermal environment which is
integrated out exactly within the path integral formulation. The approach is
numerically exact and its results reduce to known analytical findings (Marcus
theory, golden rule) in proper limits. Special attention is paid to the role of
superexchange and sequential hopping at lower temperatures in symmetric
donor-bridge-acceptor systems. In contrast to previous approximate studies,
superexchange turns out to play a significant role only for extremely high
lying bridges where the transfer is basically frozen or for extremely low
temperatures where for weaker dissipation a description in terms of rate
constants is no longer feasible. For bridges with increasing length an
algebraic decrease of the yield is found for short as well as for longer
bridges. The approach can be extended to electronic systems with more
complicated topologies including impurities and in presence of external time
dependent forces.
\end{abstract}


\maketitle

\newpage

\section{Introduction}

Electron transfer (ET) processes are one of the fundamental phenomena in
complex molecular systems \cite{jortner}, the most prominent one being the
primary step of photosynthesis \cite{marcus56,marcus85}. In the last decade or
so much theoretical and experimental efforts have been focused on bridged
molecular systems where the transfer from a donor to an acceptor is mediated by
a molecular structure connecting them
\cite{friesner,mukamel,nitzan0,ratner_nature,nitzan1,pollak}. Particular
attention has been paid to DNA complexes with to some extent still
controversial results \cite{dekker}. Of foremost importance are
donor-bridge-acceptor (DBA) systems in the strongly rising field of molecular
electronics \cite{joachim,ratner,saclay,nitzan2,hanggi}. Based on advanced
methods of synthetic chemistry, nowadays, bridge units with specific chemical
and physical properties can be produced and linked to donor/acceptor species or
contacted with metal leads and thus integrated into electrical circuits.

Since the pioneering work of Marcus and Jortner
\cite{marcus56,marcus85,jortner} the influence of residual vibronic modes and a
solvent environment on the ET is known to be extremely crucial. In fact, the
presence of a dissipative surrounding is the prerequisite for a directed
(irreversible) transport across DBA complexes meaning that bath parameters such
as temperature and spectral densities have a sensitive impact
\cite{chandler,song}. Accordingly, basically two distinct transport channels
have been identified. For high lying and short bridges quantum mechanical
tunneling gives rise to a superexchange mechanism associated with a
characteristic exponentially decreasing yield with increasing bridge length. If
the bridge is sufficiently long, however, sequential transfer processes
(hopping) dominate based on thermal activation. Theoretically the competition
between these channels has been investigated based on Redfield theory and
simple system+reservoir models in Ref.~\cite{friesner} and more recently in
Refs.~\cite{nitzan0,nitzan1,mukamel}. In these studies the mentioned
exponential length dependence as well as an algebraic decrease of the transfer
rate in case of a sequential mechanism has been found empirically. This latter
behavior has been also derived explicitly from a more rigorous formulation in
Ref.~\cite{pollak}. In most of these previous studies symmetric DBA systems
were considered with degenerate donor/acceptor and degenerate bridge states
since they already display the essential features and make the analysis
particularly transparent. Additional simplifying assumptions about the nature
of the thermal bath coupled to the ET have been invoked to obtain explicit
results since, as for most dissipative quantum systems \cite{weiss}, a
numerical description is extremely demanding.

Indeed, it has been one of the major challenges in recent years in physics and
chemistry to develop efficient algorithms to get access to a detailed analysis
of these systems. As shown first by Feynman and Vernon \cite{feynman} and in
the 1980s by Caldeira and Leggett \cite{leggett}, the path integral
representation is particularly suited as a starting point. Namely, for Gaussian
bath fluctuations the environmental degrees of freedom can be eliminated
exactly leading to the reduced system dynamics. In contrast to perturbative
approaches as e.g.~Redfield theory this formulation has the merit of being
applicable down to zero temperature, to adiabatic bath modes, and strong
friction. A direct numerical evaluation of the corresponding path integral
expression for the density matrix is not possible though. The reason for that
are the interactions, non-local in time, among forward and backward paths
generated by the bath degrees of freedom. Typically, the range of these
interactions grows with $\hbar\beta$. An efficient numerical approach that
relies on a finite interaction range for sufficiently high temperatures has
been developed by Makri and co-workers \cite{makri} and applied to a variety of
systems meanwhile.

Presently, the only numerically exact approach is given by path integral Monte
Carlo (PIMC) simulations \cite{sign-problem}. It has been successfully applied
to the spin-boson system to study e.g.~the influence of nonequilibrium initial
bath distributions \cite{lucke1}, the role of vibronic and electronic
coherences \cite{lucke1,lucke2}, and electronic transfer rates in large ranges
of parameter space \cite{mlb03}. For a DBA system with one intermediate
electronic site (three-state system) which can be seen as simple model for the
primary charge separation of photosynthesis PIMC results could describe most of
the experimental key features \cite{optimized filter}. One fundamental problem
of the PIMC, though, is intimately connected with the constitutive role of
interferences in quantum mechanical systems. Known as the {\sl dynamical sign
problem} \cite{sign-problem}, it turns out that the number of sample paths
needed to achieve a sufficient signal-to-noise ratio increases exponentially
with the simulated system time. To partially circumvent this drawback various
techniques have been introduced and shown to stabilize the simulations
considerably \cite{optimized filter,entropy,filter1,filter2,mlb98}.

The goal of this paper is twofold: On the one hand the challenge is to present
a PIMC approach applicable to larger discrete systems and longer simulation
times in order to lay the basis to treat bridge systems with more electronic
sites, a complicated topology including impurities, and/or in presence of
external periodic driving. On the other hand, motivated by the above mentioned
works on symmetric bridges, our intention is to investigate the role of the
superexchange and the sequential mechanisms in ET within a complete
system+reservoir model and with a numerically exact approach. This way, we
avoid simplifying assumptions about the bath, the system-bath coupling, or the
structure of the bath along the molecular bridge. Accordingly, our results can
directly be compared with analytical rate expressions derived within Marcus
theory \cite{weiss}, golden rule approaches \cite{GR2}, or cluster expansion
methods \cite{nica}. Surprisingly, such an analysis of numerical results within
the context of the available theoretical findings is still missing.

The paper is organized as follows: We start in Sec.~\ref{Sec: Theory} by
defining the system and the bath and introducing its path integral
representation. In Sec.~\ref{Sec: Population dynamics} the connection to the
description in terms of Master equations is revealed and known results for the
electronic transfer rates are discussed. Then, in Sec.~\ref{Sec: Simulation
method} the PIMC used and developed in this paper is presented, which leads to
a substantial improvement in efficiency compared to previous PIMC schemes. Only
this allows us to study in Sec.~\ref{Sec: Numerical results} first the
three-state system and then complexes with longer bridges in detail,
particularly with very good signal to noise ratio and larger times. A
Conclusion contains our main results and some remarks about subsequent work.

\section{Theory}
\label{Sec: Theory}
\subsection{Dynamics of the dissipative $d$-level system}
\label{Subsec: Dynamics of the dissipative d-level system}

\begin{figure}
\epsfxsize=8cm
\epsffile{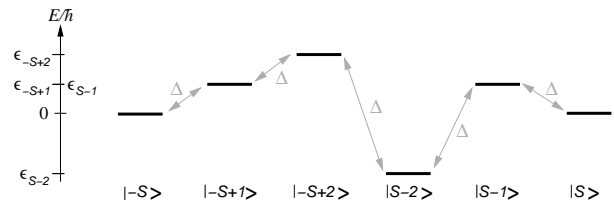}
\caption[]{\label{fig_1} Molecular chain for $S=2^{ 1\!}/_2$ $(d=6)$ and
  constant nearest neighbor coupling $\Delta$.}
\end{figure}

We investigate the dynamics of a single electron moving on a chain of $d=2 S+1$
discrete sites, labeled by a discrete variable $-S \leq s \leq S$ with spacing
1, separated by equal distances $a$ but arbitrary energies $\hbar\epsilon_s$
(see Fig.~\ref{fig_1}). Electronic motion is facilitated through tunneling
between sites $s$ and $s'$, with a real tunneling amplitude
$\Delta_{s,s'}$. The electronic coordinate can then be expressed as
\begin{equation} \label{path}
q(t) = a \cdot s(t) \;,
\end{equation}
where $-S \le s(t) \le S$. The position operator thus is equivalent to the spin
$S$ operator
\begin{equation}
a{\mathbf S}_z|s\rangle = as|s\rangle \;,
\end{equation}
with $|s\rangle$ denoting the (orthonormal) localized electronic states. In
terms of electron transfer, $|{-S}\rangle$ and $|S\rangle$ represent the donor
and acceptor, respectively, while the other states are referred to as the
bridge states.

The free $d$-level system ($d$LS) Hamiltonian can then be written as
\begin{equation} \label{H_dLS-Hamiltonian}
H_{d\rm LS} = \hbar {\mathbf E}_z - \hbar {\mathbf S}_x \;,
\end{equation}
where ${\mathbf E}_z$ describes the energetic distribution of the electronic
sites according to
\begin{equation}
{\mathbf E}_z|s\rangle = \epsilon_s|s\rangle \;,
\end{equation}
while ${\mathbf S}_x$ governs the tunneling,
\begin{equation} \label{sx}
{\mathbf S}_x=\sum_{s<s'=-S}^{S}\, {\Delta}_{s,s'} \, \left(\,
|s\rangle\!\langle s'| + |s'\rangle\!\langle s|\, \right) \;.
\end{equation}
Of particular importance is the case of nearest neighbor tunneling only with
\begin{eqnarray} \label{nncoupling}
{\mathbf S}_x|s\rangle &=& \Delta_{s-1,s}|s-1\rangle + \Delta_{s,s+1}
|s+1\rangle \nonumber\\
&& {\rm for}\ {-S} < s < S\;, \nonumber\\
{\mathbf S}_x|{-S}\rangle &=& \Delta_{-S,-S+1}|{-S}+1\rangle\;, \nonumber\\
{\mathbf S}_x|S\rangle &=& \Delta_{S,S+1}|S-1\rangle\;.
\end{eqnarray}
We already note here that the propagation of the free $d$-level system is most
conveniently performed not in the site representation used above, but rather in
the eigenstate representation of $H_{d\rm LS}$ (for details see Sec.~\ref{Sec:
Simulation method}). This is completely equivalent to the transformation
between (localized) Wannier states and delocalized molecular orbitals (Bloch
states) in infinite tight binding lattices. The site representation is
particularly suited, however, to include the interaction with solvent and
vibronic degrees of freedom which later on will be integrated out. For this
purpose, we embed the $d$LS in a system-plus-reservoir model, leading to the
total Hamiltonian \cite{weiss}
\begin{eqnarray} \label{hamilton}
H &=& H_{d\rm LS} + H_I + H_B \nonumber\\
&=& H_{d\rm LS}- a{\mathbf S}_z \sum_\alpha c_\alpha X_\alpha
 + (a{\mathbf S}_z)^2 \sum_\alpha {c_\alpha^2 \over 2m_\alpha \omega_\alpha^2}
\nonumber\\
&&{} + \sum_\alpha \left( {P_\alpha^2 \over 2m_\alpha}
 + {1\over2}m_\alpha\omega_\alpha^2 X_\alpha^2 \right) \; .
\end{eqnarray}
The residual degrees of freedom are thus modeled by an infinite collection of
harmonic oscillators, $H_B$, which bilinearly couple to the position of the
electron ($H_I$). The so-called counterterm quadratic in the $d$LS position
operator $a{\mathbf S}_z$, which reduces to a physically irrelevant constant
for $S=1/2$, prevents a renormalization of the electronic energy levels by the
oscillator bath \cite{weiss}. As discussed in detail in
Refs.~\cite{chandler,song,weiss,GR2} this provides a reasonably accurate
description of reality for the great majority of ET systems. It turns out that
for the electronic dynamics the environmental parameters enter via the spectral
density
\begin{equation}
J(\omega) = \frac{\pi a^2}{2\hbar} \sum_\alpha \frac{c_\alpha^2}
 {m_\alpha\omega_\alpha} \delta(\omega-\omega_\alpha) \;,
\end{equation}
which effectively becomes a continuous function of $\omega$ for a
condensed-phase environment. The Gaussian statistics of the environment is
determined by the complex-valued bath autocorrelation function which for real
time $t$ reads
\begin{eqnarray} \label{lz}
L(t) &=&\frac{1}{\hbar^2}\left\langle \left(\sum_\alpha c_\alpha
X_\alpha(t)\right)\left(\sum_\alpha c_\alpha X_\alpha(0)\right) \right\rangle_\beta\nonumber\\
&=& {1\over\pi}\int_0^\infty \!d\omega\, J(\omega)
{\cosh[\omega(\hbar\beta/2-it)] \over \sinh(\hbar\beta\omega/2)} \;,
\end{eqnarray}
where $\beta=1/k_B T$. For $S=1/2$, Eq.~(\ref{hamilton}) represents the
celebrated {\sl spin-boson} (SB) model, with numerous applications in
solid-state physics \cite{weiss}. In the special case of a constant nearest
coupling, i.e.~$\Delta_{s,s'}=\delta_{s'-s,1}\Delta$ for all $s,s'$, the
described compound reduces to a generalization of the SB model, as it in fact
then resembles a ``spin-$S$-boson model'' \cite{weiss}. In the following, we
repeatedly will take advantage of the striking similarity. To make connection
to (classical) two-state ET theory, we also mention the classical {\sl
reorganization energy} $\hbar \Lambda_{\rm cl}$ of the Marcus theory
\cite{marcus56,marcus85},
\begin{equation} \label{reorg}
\Lambda_{\rm cl} = {1\over\pi} \int_0^\infty\!d\omega\, {J(\omega) \over
\omega} \; .
\end{equation}

For the dissipative electronic dynamics we focus on the {\it time-dependent
site populations},
\begin{equation} \label{populations}
P_{s_f,s_i}(t) = {\rm Tr}\left\{e^{iHt/\hbar} |s_f\rangle\!\langle s_f|
e^{-iHt/\hbar} W_i(0)\right\} \;,
\end{equation}
which are normalized, $\sum_{s_f=-S}^S P_{s_f,s_i}(t) = 1$, and where $W_i(0)$
specifies the initial state of the total compound. In most theoretical and
experimental works an initial separation of the electron and the environment is
assumed corresponding to an initial density matrix
\begin{equation} \label{initial_perp_P}
W_i(0) = Z_B^{-1} |s_i\rangle\!\langle s_i| e^{-\beta(H_B-s_i \mu{\cal
E})} \;,
\end{equation}
with the electron held fixed in state $|s_i\rangle$ {and the bath normalization
$Z_B = {\rm Tr}\{e^{-\beta H_B}\}$ assuring} {the full system's density matrix
to be normalized for all times}. For a transfer process across the entire chain
one typically prepares the electron initially in the donor state, i.e.\
$|s_i\rangle=|{-S}\rangle$ \cite{friesner,nitzan1}. Above, $\mu$ is the dipole
moment of the electron, and ${\cal E}$ denotes the dynamical polarization of
the bath \cite{weiss}, which is equilibrated with respect to the initial
position of the electron. By comparing with Eq.~(\ref{hamilton}), we see that
$\mu {\cal E} = a\sum_\alpha c_\alpha X_\alpha$. As pointed out in
Ref.~\cite{mlb03}, this ``standard preparation'' often used in ET experiments
is especially suitable for a theoretical description of thermal transfer rates.

The present formulation can easily be extended in order to describe the impact
of an external time dependent driving force. This as well as abandoning the
restriction to nearest neighbor coupling will be the subject of a subsequent
paper.

\subsection{Path integral representation}
\label{Subsec: Path integral representation}

The path integral representation provides a formally exact expression for the
dynamics of the electronic population and is thus the starting point for a
numerically exact Monte-Carlo (MC) scheme. The standard procedure is then to
eliminate the bath degrees of freedom to arrive at the reduced dynamics. This
can be done exactly due to the harmonic nature of the bath. As shown in
Ref.~\cite{weiss}, one thus obtains for Eq.~(\ref{populations})
\begin{equation} \label{pops_path}
P_{s_f,s_i}(t) = \frac{1}{Z} \oint\!{\cal D}\tilde{s}\;
 \delta_{\tilde{s}(t),s_f} \exp\left\{ {i\over\hbar}S_{d\rm LS}[\tilde{s}] -
 \Phi[\tilde{s}] \right\} \;.
\end{equation}
Here the path integration runs over closed paths $\tilde{s}(\tilde{t})$
starting at $\tilde{s}(0)=s_i$ and propagating along the real-time contour
$\tilde{t} \in 0 \rightarrow t \rightarrow 0$, which connect the forward and
backward paths $s(t')$ and $s'(t')$, respectively. Furthermore, $S_{d\rm
LS}[s]$ denotes the total action of the free $d$LS, and the influence of the
traced-out bath is encoded in the {\sl Feynman-Vernon influence functional}
$\Phi[s]$ \cite{feynman}. In terms of the bath autocorrelation function
(\ref{lz}), it reads \cite{weiss}
\begin{eqnarray} \label{influence-exponent}
\Phi[s, s'] &=& \int_0^{t}\!dt'\! \int_0^{t'}\!dt'' [s(t') - s'(t')]
\ [{L}(t'-t'')s(t'') \nonumber\\
&& \qquad {}- {L}^\ast(t'-t'')s'(t'')] \nonumber\\
&& {}+ i{{\hat{\mu}} \over 2}\int_0^{t}\!dt' [s^2(t') - s'^2(t')] \;,
\end{eqnarray}
where
\begin{equation} \label{muu}
{\hat{\mu}} = {2 \over \pi} \int_0^\infty\!d\omega {J(\omega) \over
\omega} = 2\Lambda_{\rm cl} \;.
\end{equation}
The influence functional introduces long-ranged nonlocal interactions among the
$d$LS paths so that in general an explicit evaluation of the remaining path
integral in Eq.~(\ref{pops_path}) is possible only numerically. Before we
explain the details of the PIMC scheme in Sec.~\ref{Sec: Simulation method},
however, we focus in the next Section on the description of the population
dynamics in terms of Master equations with associated transition
rates. Corresponding results will provide the appropriate tools in order to
analyze and understand the MC data.

\section{Population dynamics}
\label{Sec: Population dynamics}
\subsection{Master equations}
\label{Subsec: Master equations}

As has been derived in Ref.~\cite{nica}, the dynamics of the population for a
spin-$S$-boson system can be cast exactly into the form of a retarded non-local
(in time) Master equation
\begin{equation} \label{exactME}
\frac{dP_{s_f,s_i}(t)}{dt}=\sum_{s=-S}^S \int_0^t dt'\
\tilde{\Gamma}_{s,s_i}(t-t')\, P_{s,s_i}(t')
\end{equation}
with time dependent transition rates $\tilde{\Gamma}_{s,s_k}(t)$ the Laplace
transforms of which are given by appropriate cluster functions. Provided the
dissipative influence is sufficiently strong or temperature sufficiently high,
the electronic dynamics (\ref{exactME}) can be simplified to a conventional
Master equation. In essence, one invokes a coarse graining procedure in time so
that after some initial transient motion, i.e.~for $t > \tau_{\rm trans}$, the
population dynamics obeys
\begin{equation} \label{convME}
\dot{P}_{s_f,s_i}(t)=\sum_{s=-S}^S \Gamma_{s,s_i}\, P_{s,s_i}(t)\; .
\end{equation}
Here, the nearest neighbor transition rates $\Gamma_{s,s\pm 1}$ are related to
sequential hopping processes, while $\Gamma_{-S,S}$ describes direct transfer
from donor to acceptor without any real populations in the bridge states
(superexchange).

In the case studied here in more detail, namely, degenerate donor and acceptor
energies ($\epsilon_{-S}=\epsilon_S=0$), degenerate bridge energies
($\epsilon_{-S+1}=\ldots=\epsilon_{S-1}=\epsilon$), and nearest neighbor
tunneling with $\Delta_{s,s'}=\delta_{s'-s,1}\Delta$, the number of independent
rates shrinks considerably. Eq.~(\ref{convME}) becomes particularly simple in
case of purely sequential transport where it can be written as
\begin{equation} \label{bridgeME}
\dot{{\mathbf P}} = A {\mathbf P} \;,
\end{equation}
with the rate matrix
\begin{widetext}
\begin{equation} \label{matrix}
A = \left(
 \begin{array}{ccccccccc}
 -\Gamma_{DB} & \Gamma_{BD} && 0 && \multicolumn{3}{c}{\cdots} & 0\\
 \Gamma_{DB} & -(\Gamma_{BD} + \Gamma_B) && \Gamma_B && 0 && \cdots & \\
 0 & \Gamma_B && -2\Gamma_B && \Gamma_B && 0 & \\
 \vdots & & \ddots && \ddots && \ddots && \vdots \\
 & 0 && \Gamma_B && -2\Gamma_B && \Gamma_B & 0 \\
   & \cdots && 0 && \Gamma_B && -(\Gamma_B + \Gamma_{BA}) & \Gamma_{AB} \\
 0 & \multicolumn{3}{c}{\cdots} && 0 && \Gamma_{BA} & -\Gamma_{AB} \\
 \end{array}
 \right)
\end{equation}
\end{widetext}
Here, ${\mathbf P}=(P_D,P_{B1},\ldots,P_{Bb},P_A)$ where $P_D$ ($P_A$) refers
to the donor (acceptor) population $P_{-S}$ ($P_S$) and $P_{Bi}$ to the
population $P_{i-S}$ of the $b=d-2$ bridge states. The rate constants
$\Gamma_{DB}$ ($\Gamma_{BD}$) govern the sequential forward (backward)
transport between donor and first bridge state, $\Gamma_{AB}$ ($\Gamma_{AD}$)
the sequential forward (backward) transport between acceptor and last bridge
state, and $\Gamma_B$ the sequential transport on the bridge. Coherent
transport (superexchange) would be accounted for by additional rate
contributions in the upper and lower triangular of $A$.

According to Eq.~(\ref{bridgeME}) the populations $P_s(t)$ reflect
multi-exponential dynamics, with the corresponding rates given by the
eigenvalues of the rate matrix $A$. The approach to equilibrium happens to be
on the relaxation time scale $\tau_R$ which is the inverse of the smallest of
these eigenvalues $\Gamma_R$. Forward and backward rates in $A$ are connected
via detailed balance,
\begin{equation}
\Gamma_{BD} = {P_B^\infty \over P_D^\infty} \Gamma_{DB}
 = {2P_B^\infty \over 1-b P_B^\infty} \Gamma_{DB} \;,
\end{equation}
where $P_B^\infty$ and $P_D^\infty$ denote the equilibrium occupation
probabilities of the bridge states and the donor, while for symmetry reasons
$P_A^\infty = P_D^\infty$, $\Gamma_{BA} = \Gamma_{BD}$, and $\Gamma_{AB} =
\Gamma_{DB}$. The equilibrium occupation probabilities are Boltzmann
distributed with respect to an effective electronic parameter
$\tilde{\epsilon}(\beta)$ which, for strong damping and/or higher temperatures,
converges with the physical bias $\epsilon$ \cite{hermann_talkner}. It is then
straightforward to calculate $P_s(t)$ from Eq.~(\ref{bridgeME}) or the more
general expression (\ref{convME}) which in the appropriate parameter range
gives an accurate description of the exact dynamics for times $t$ larger than
some transient time scale $t_{\rm trans}$. The corresponding expressions for $d
= 3$ are given in App.~\ref{App: Expressions for the populations}.

Hence, once MC data for the populations are generated, the associated rate
constants are obtained by fitting the electronic dynamics obtained from
Eqs.~(\ref{bridgeME}) and (\ref{matrix}) to the numerical data. At sufficiently
low temperatures, however, coherent oscillations govern the population dynamics
over a long period of time and the transfer can no longer be characterized by
time independent decay rates. Outside this range information about the impact
of superexchange can be extracted by comparing fits to the sequential
(\ref{bridgeME}) and the complete model (\ref{convME}).

\subsection{Sequential and superexchange rate expressions}
\label{rate}

As already noted, basically two transport channels have been found in previous
studies on bridge mediated ET, namely, a sequential channel and a superexchange
channel. The latter one is expected to dominate for high lying and sufficiently
short bridges, whereas in other cases the former one prevails. Here we briefly
collect available analytical results for the transitions rates,
particularly for symmetric DBA systems (degenerate donor and acceptor energies,
degenerate bridge states), to analyze our PIMC results accordingly (see
Sec.~\ref{Sec: Numerical results}).

In the classical limit $\hbar\beta\omega_c \ll 1$ the sequential forward rate
is given by the well known Marcus result \cite{marcus85}
\begin{equation} \label{marcus}
\Gamma_{DB, \rm Marcus} = \frac{\Delta^2}{1+\pi \Delta^2/(\Lambda_{\rm
cl} \omega_c)}
 \sqrt{{\pi \hbar \over \Lambda_{\rm cl} k_B T}}\ e^{-\beta F^\ast(\epsilon)}
\end{equation}
with the activation free energy barrier $F^\ast(\epsilon) = \hbar(\epsilon +
\Lambda_{\rm cl})^2/(4\Lambda_{\rm cl})$ where $\epsilon=\epsilon_B-\epsilon_A$
is the energy gap between bridge and donor (acceptor). As long as a rate
description is valid at all, this expression is applicable from the weak
adiabatic domain ($\omega_c<\Delta$), where it eventually becomes independent
of $\Delta$, to the nonadiabatic range ($\omega_c\gg \Delta$). Recently, for
adiabatic to moderate nonadiabatic environments an extension of this rate
expression to lower temperatures where quantum fluctuations and nuclear
tunneling come into play has been derived in Ref.~\cite{lehle}.

In the quantum domain $\hbar\beta\omega_c{\textstyle {\lower 2pt \hbox{$>$}
\atop \raise 1pt \hbox{$ \sim$}}}1$ and for small tunneling amplitudes $\Delta
\ll \omega_c$ (nonadiabatic range) transfer rates are calculated perturbatively
by invoking Fermi's golden rule. For the sequential forward rate the result is
\cite{GR2,weiss}
\begin{equation} \label{grs}
\Gamma_{DB, \rm GR}(\epsilon)=\Delta^2 \int_{-\infty}^{\infty} dt
\exp[-i\epsilon t- Q(t)]
\end{equation}
with the damping kernel $Q(t)$ specified in Eq.~(\ref{Q1}). The golden rule
rate expression reduces for high temperatures to the nonadiabatic limit of the
Marcus formula (\ref{marcus}) and captures nuclear tunneling at low $T$. Within
a perturbative treatment in powers of $\Delta^{2n}$, superexchange rates appear
as higher order contributions $n=2,3,4,\ldots$. In case of a three-state system
with $\epsilon_D=\epsilon_A$, $\Gamma_{-1,1}$ is obtained in leading order
$n=2$ as \cite{nica}
\begin{eqnarray} \label{super1}
\Gamma_{S, \rm GR}&=& 2 {\Delta^4}\, {\rm Re}\bigg\{\int_0^\infty d\tau_1
d\tau_2d\tau_3
\ \exp[Q^\ast(\tau_3) \nonumber\\
&&\quad{} - Q(\tau_2+\tau_3) - Q(\tau_1+\tau_2+\tau_3) - Q(\tau_2)\nonumber\\
&&\quad{} - Q(\tau_1+\tau_2)+Q(\tau_1)]\nonumber\\
&&\times \exp[-i \epsilon (\tau_1-\tau_3)]\bigg\} \;,
\end{eqnarray}
Even for this more involved expression simplifications arise in certain
limits. In the high temperature range $\hbar\beta\omega_c \ll 1$, where
$Q(\tau)$ can be expanded for short times, one obtains the classical
superexchange rate \cite{nica}
\begin{eqnarray} \label{super2}
\Gamma_{SCL, \rm GR} &=& \frac{\Delta^4}{2 (\epsilon-\Lambda_{\rm
cl})^2}\
\sqrt{\frac{\pi \hbar\beta}{\Lambda_{\rm cl}}}\nonumber\\
&&\times \left\{{\rm
 e}^{-\beta\hbar\Lambda_{\rm cl}}-\frac{1}{2}\left[{\rm e}^{-\beta F_+}+{\rm
 e}^{-\beta F_-}\right]\right\}
\end{eqnarray}
with the activation energies $F_+=\hbar(\epsilon+\Lambda_{\rm
cl})^2/4\Lambda_{\rm cl}$ and $F_-=\hbar(\epsilon-3\Lambda_{\rm
cl})^2/4\Lambda_{\rm cl}$. This rate is well-defined even in the resonant case
$\epsilon\to \Lambda_{\rm cl}$ where, however, also the incoherent transfer is
extremely efficient with a strong bridge population.

In presence of a quantum bath $\hbar\beta\omega_c {\textstyle {\lower 2pt
\hbox{$>$} \atop \raise 1pt \hbox{$ \sim$}}}1$ the classical reorganization
energy drops out of the golden rule rate, see Eq.~(\ref{super2}), since nuclear
tunneling is relevant. The superexchange expression can be simplified for a
high-lying intermediate state where the DBA system can effectively be treated
as an AD system with effective coupling \cite{nica}, i.e.,
\begin{equation} \label{super3}
\Gamma_{SQM, \rm GR}\approx \frac{\Delta^4}{\epsilon^2}\,
\int_{-\infty}^\infty d\tau \exp[-4 Q(\tau)]\; .
\end{equation}

\section{Simulation method}
\label{Sec: Simulation method}

As already addressed, the time evolution of a dissipative quantum system can in
general not be evaluated analytically. In the
past the PIMC method has been proven as a very promising approach to obtain numerically exact results even in
regions of parameter space where other approximative methods fail. In
particular, in recent experiments on molecular wires contacted with metal
junctions \cite{weber} low temperature measurements down to 30K have been
performed which necessitates the inclusion of strong quantum effects also in
the environment.

For this purpose we turn to the path integral representation (\ref{pops_path})
and employ a discretization described in detail in Refs.~\cite{optimized
filter,mlb03}. For the forward and backward paths the time axis is sliced via
$q$ uniformly spaced points with discretization steps $\tau = t/q$. The path
integral in Eq.~(\ref{pops_path}) then becomes
\begin{equation} \label{pop-discretized}
P_{s_f, s_i}(t) = \sum_{\{s_j\}} \delta_{s_{q+1},s_f} \rho[\{s_j\}]
\end{equation}
with
\begin{equation} \label{density}
\rho[\{s_j\}] = \left[\prod_{k=1}^{2 q}
K(s_k,s_{k+1})\right]e^{-\Phi[\{s_j\}]} \;.
\end{equation}
The sum runs over all realizations of the discretized spin path $\{s_j\} =
\{s_1\equiv s_i,s_2,\dots,s_{2q}, s_{2q+1}\equiv s_i\}$, and $K(s_j,s_{j+1})$
denotes the coordinate representation of the free $d$LS propagation over the
time interval $\tau$, i.e.,
\begin{equation}
K(s,s',\tau)=\langle s|\exp(-i\tau H_{d\rm LS}/\hbar)|s'\rangle \;.
\end{equation}
While in principle for the $d$LS Hamiltonian this propagator can be obtained by
exploiting the symmetric Trotter splitting $\exp(-i\tau H_{d\rm LS}/\hbar) =
\exp(-i\tau{\mathbf E}_z/2) \exp(i\tau {\mathbf S}_x) \exp(-i\tau{\mathbf
E}_z/2) + O(\tau^3)$ and evaluating $\langle s_{j+1}|\exp(i\tau {\mathbf
S}_x)|s_j\rangle$, a more accurate approach follows from the eigenstate
representation
\begin{equation} \label{syseigen}
H_{d\rm LS} |\phi_N\rangle = E_N|\phi_N\rangle\,, \quad N=1,\ldots, d\;.
\end{equation}
This way, one obtains
\begin{equation} \label{sysprop2}
K(s,s',\tau) = \sum_{N=1}^d \langle s|\phi_N\rangle \langle\phi_N|s'\rangle\
{\rm e}^{-i \tau E_N/\hbar} \;,
\end{equation}
which can be easily computed numerically once the $d$LS's parameters are
specified. Another advantage of this formulation is that it can be immediately
applied to all quantum systems which can be mapped effectively onto a $d$LS,
e.g.~by a proper reduction of its Hilbert space.

To arrive at a discretized form of the influence functional
(\ref{influence-exponent}), the sum and difference coordinates
\begin{equation} \label{sum and difference coordinates}
\eta(t') \equiv s(t') + s'(t') \;, \qquad \xi(t) \equiv s(t') - s'(t')
\end{equation}
are introduced, which read $\eta(t') = \eta_j$ ($\xi(t') = \xi_j$) for $t' \in
[(j-1)\tau - \tau/2, (j-1)\tau + \tau/2]$ in their discretized form. The sum
paths are also considered as ``quasi-classical'', while the difference paths
capture quantum fluctuations \cite{optimized
filter,weiss}. Equation~(\ref{influence-exponent}) finally can be written as
\begin{eqnarray} \label{influence}
\Phi[s_i, \eta, \xi] &=&
i\sum_{j=2}^q \xi_j \hat{X}^{(s_i)}_j\\
&&{}+ \sum_{j \ge k=2}^q \xi_j (i X_{j-k} \eta_k +
\Lambda_{j-k} \xi_k) \nonumber
\end{eqnarray}
with
\begin{eqnarray}
\hat{X}^{(s_i)}_j &=& 2s_i\, {\rm Im}\{Q((j-2)\tau) -
Q((j-1)\tau)\}\nonumber\\
&& {\rm for}\ 2
\le j \le q \;,\nonumber\\
\Lambda_0 &=& {\rm Re}\{Q(\tau)\} \;,\nonumber\\
X_0 &=& {\rm Im}\{Q(\tau)\} \;,\nonumber\\
\Lambda_l &=& {\rm Re}\{Q((l-1)\tau) +
Q((l+1)\tau) - 2Q(l\tau)\} \nonumber\\
&&{\rm for}\ 1 \le l \le q-2 \;,\nonumber\\
X_l &=& {\rm Im}\{Q((l-1)\tau) +
Q((l+1)\tau) - 2Q(l\tau)\} \nonumber\\
&&{\rm for}\ 1 \le l \le q-2 \;.
\end{eqnarray}
Here, one has introduced the twice-integrated bath autocorrelation function
$Q(t)$ defined by $\ddot{Q}(t) = L(t)$, $Q(0) = 0$ with $\dot{Q}(0) =
i\Lambda_{\rm cl}$. Explicitly it is gained from Eq.~(\ref{lz}) to read
\begin{eqnarray} \label{Q1}
Q(t) &=& {1\over\pi} \int_{0}^{\infty}\!d\omega {J(\omega)\over\omega^2}
\{\coth(\hbar\beta\omega/2))[1-\cos(\omega t)] \nonumber\\
&& \qquad{} + i\sin(\omega t)\}\; .
\end{eqnarray}
In the sequel we consider a spectral density of the form
\begin{equation} \label{spectral}
J(\omega)= 2 \pi \alpha \omega {\rm e}^{-\omega/\omega_c} \;,
\end{equation}
which is equivalent to ohmic damping with a cut-off frequency $\omega_c$. In
this case $Q(t)$ can be calculated analytically and one obtains
\begin{equation}
Q(t) = 2\alpha\Bigg[ \ln(1+i\omega_c t)
 - \ln\frac{\Gamma(\Omega+i t/\hbar\beta)\Gamma(\Omega-it/\hbar\beta)}
 {\Gamma^2(\Omega)} \Bigg]
\end{equation}
with $\Omega = 1 + 1/(\hbar\beta\omega_c)$ and the Gamma function $\Gamma(z)$.

Equations~(\ref{pop-discretized}), (\ref{density}) and (\ref{influence})
constitute a discretized form of the populations (\ref{populations}) and thus
provide a starting point for PIMC simulations. Unfortunately, this method is
handicapped by the dynamical sign problem \cite{sign-problem}. It originates
from quantum interferences between different system paths $\{s_j\}$, causing a
small signal-to-noise ratio of the stochastic averaging procedure: The
exponential increase of the paths' configuration space with the maximum real
time under study results in an exponential decrease of the signal-to-noise
ratio. Various procedures to mitigate the sign problem have been proposed in
the past, like maximum entropy methods \cite{entropy}, filter techniques
\cite{filter1,filter2} or the multilevel blocking approach
\cite{mlb98,mlb03}. Here we employ a filter technique optimized for dissipative
spin systems as suggested by Egger and Mak \cite{optimized filter} which
exploits the special symmetries of the influence functional.

The backbone of this approach is the observation that the absence of a
self-energy like term for the quasi-classical paths $\{\eta_j\}$ in
Eq.~(\ref{influence}) allows to express the summation over the latter as a
series of $q-1$ matrix multiplications. This usually requires drastically less
computational effort than performing the corresponding sums in
Eq.~(\ref{pop-discretized}) and thus can be carried out explicitly. It reduces
the degrees of freedom from the $2q-1$ variables $\{\eta_{2\le j \le q+1},
\xi_{2\le j \le q}\}$ to the $q-1$ quantum variables $\{\xi_{2\le j \le q}\}$
and therefore significantly improves the numerical stability of the
corresponding MC simulations. Explicitely, after switching to the
quasi-classical and quantum coordinates (\ref{sum and difference coordinates})
and defining the $[2(S-|\xi_j|)+1]\times[2(S-|\xi_{j+1}|)+1]$ matrices
$\hat{K}^{(j)}$ according to
\begin{eqnarray} \label{new free propagator}
\lefteqn{\hat{K}^{(j)}_{\eta_j,\eta_{j+1}}(\xi_j,\dots,\xi_q)} \nonumber\\
&& \equiv \exp\!\left(
-i\eta_{j+1}\sum_{k>j} \xi_k X_{k-j-1} \right) \nonumber\\
&& \times K^{}((\eta_j+\xi_j)/2, (\eta_{j+1}+\xi_{j+1})/2) \nonumber\\
&& \times K^\ast((\eta_j-\xi_j)/2, (\eta_{j+1}-\xi_{j+1})/2) \;,
\end{eqnarray}
the population (\ref{pop-discretized}) becomes
\begin{eqnarray} \label{population}
P_{s_f, s_i}(t) &=& \sum_{\{\xi_j\}, \eta_{q+1}} \delta_{\eta_{q+1},2s_f}
 \exp\!\left[-\sum_{j=2}^q \xi_j \Bigg(i\hat{X}^{(s_i)}_j \right. \nonumber\\
&&\left.\left.{} +\sum_{k \ge j} \xi_k \Lambda_{k-j} \right)\right]
 \nonumber\\
&&\quad\times
\hat{K}^{(1)}(0, \xi_2, \dots, \xi_q) \cdots \hat{K}^{(q)}(\xi_q) \;.
\end{eqnarray}
Here, the sum over the $\{\eta_{j \le q}\}$ has been written as a product over
the $q$ matrices $\hat{K}^{(j)}$, where the first (last) of them is only a
row (column) vector due to $\eta_1 \equiv 2s_i$ ($\eta_{q+1} \equiv
2s_f$). This matrix product can be performed explicitely such that only the sum
over the quantum variables $\{\xi_j\}_{2 \le j \le q}$ has to be evaluated by
PIMC simulations. The resulting mitigation of the sign problem allows for
numerically stable simulations which consume significantly less CPU time than
corresponding ones evaluating the standard expression
(\ref{pop-discretized}). Moreover, another considerable speedup can be gained
by optimizing the MC weight with respect to the dissipative regime (see
App.~\ref{App: Computational details}).

We note in passing that the above described approach also resembles the
multilevel blocking strategy \cite{makegger}: On the first level, the harmonic
bath degrees of freedom are integrated out, while on the second level the
quasi-classical coordinates get eliminated; the addend in
Eq.~(\ref{population}) then corresponds to the respective level-2 bonds. It
furthermore seems noteworthy that, while here we only present results for
electronic systems with constant nearest-neighbor coupling, $\Delta_{s,s'} =
\delta_{s'-s,1}\Delta$, the above outlined approach is suitable for arbitrary
electronic systems as long as they can be described by a finite Hilbert
space. Corresponding applications will be shown in a subsequent paper.

\section{Numerical results}
\label{Sec: Numerical results}

Next we present results for a symmetric donor-bridge-acceptor (DBA) system
(degenerate donor and acceptor energies, degenerate bridge states with energy
gap $\hbar\epsilon$ between bridge and donor/acceptor) with constant nearest
neighbor coupling obtained from PIMC simulations as described above. Albeit its
simplicity this model captures essential features of bridge mediated ET and can
serve as a basis for more elaborate studies. In part, this work was motivated
by former theoretical approaches on symmetric DBA systems
\cite{friesner,nitzan1,pollak} which, however, relied on simpler
system+reservoir models and/or approximate numerical methods. In particular, in
contrast to these latter, the formulation (\ref{hamilton}) takes the dynamics
of the vibronic structure of the DBA compound fully into account. Further, the
MC approach is not plagued by the limitations of Redfield theory, but rather
applies also to low temperatures, slower bath modes, and stronger
dissipation. Hence, it reproduces in the proper limits the known analytical
findings specified in Sec.~\ref{rate}.

Clearly, the ET process considered here eventually leads to thermal
equilibration of the electronic sites.
Due to the relation between relaxation rates and conductances in stationary
non-equilibrium \cite{nitzan2}, our results give also insight, at least
qualitatively, into ET through metal-molecule-metal junctions. In fact, to get
quantitative results, the present approach can in principle be extended to the
latter case by eliminating the electronic states in the metal contacts as an
additional bath. Corresponding work is in progress.

The superexchange mechanism is a truly quantum mechanical coherence
phenomenon. Thus, in order to investigate this scenario in more detail within
the MC approach, we choose parameters such that for lower temperatures the
population dynamics exhibits coherent oscillations, while for slightly higher
$T$, with a bath that is still quantum mechanically, a rate description is
applicable. We take a damping coefficient $\alpha = 1/4$ well below the
coherent-incoherent transition $\alpha = 1/2$ \cite{weiss} and a moderate
cut-off frequency $\omega_c/\Delta = 5$ so that adiabatic effects are expected
to show up. Note that the cutoff frequency $\omega_c$ also defines the maximum
of the spectral density distribution (\ref{spectral}), such that even
frequencies somewhat larger than $\omega_c$ contribute. The classical
reorganization energy follows as $\Lambda_{\rm cl} = 2\alpha\omega_c =
2.5\Delta$. Accordingly, we expect superexchange, if important, to be
particularly pronounced.

\begin{figure}
\epsfxsize=8cm
\epsffile{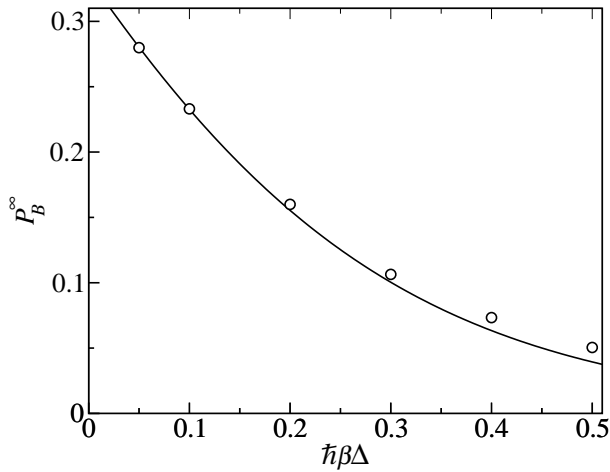}
\caption[]{\label{fig_2} Equilibrium occupation probabilities $P^\infty_B$ of
  the bridge state for $S = 1$, $\epsilon/\Delta = 5$, $\alpha = 1/4$ and
  $\omega_c/\Delta = 5$ as a function of inverse temperature. Circles denote
  imaginary-time QMC results (with error bars smaller than symbol size) while
  the solid line refers to $1/[2e^{\hbar\beta\epsilon}+1]$ according to a
  Boltzmann distribution.}
\end{figure}

\begin{figure}
\epsfxsize=8cm
\epsffile{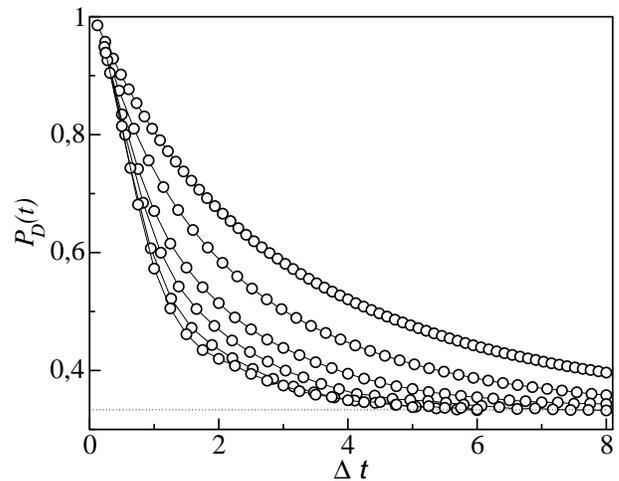}\\[0.8cm]
\epsfxsize=8cm
\epsffile{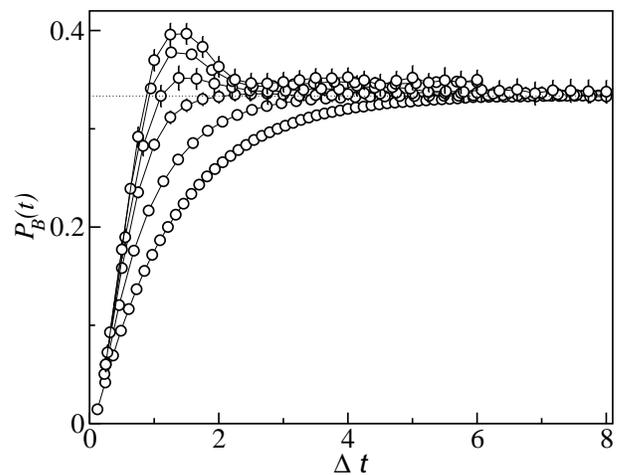}\\[0.8cm]
\epsfxsize=8cm
\epsffile{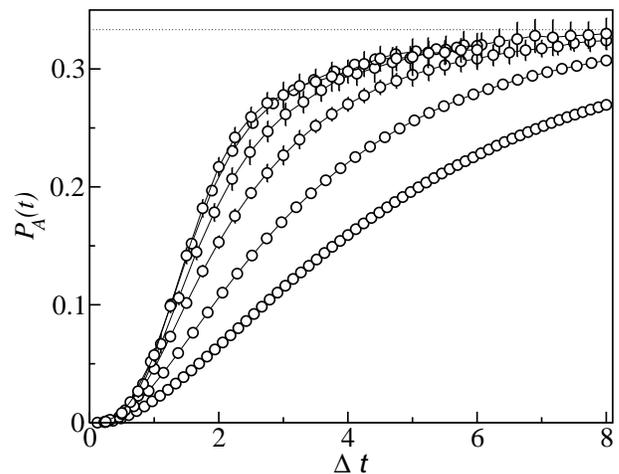}
\caption[]{\label{fig_3} Donor, bridge, and acceptor populations $P_D(t)$,
 $P_B(t)$, and $P_A(t)$, respectively (from top to bottom), for $S = 1$,
 $\epsilon/\Delta = 0$, $\alpha = 1/4$, $\omega_c/\Delta = 5$, and (top to
 bottom for donor, bottom to top for bridge and acceptor) $\hbar\beta\Delta$ =
 0.05, 0.1, 0.2, 0.3, 0.4, and 0.5. Error bars correspond to one
 standard deviation in each direction. Dotted lines denote the equilibrium
 population $P_D^\infty = P_B^\infty = P_A^\infty = 1/3$. The solid lines are
 guides for the eye only.}
\end{figure}

To confirm the proper choice of the above parameters and to fix the temperature
range where a description in terms of rate constants applies, we start by
presenting results for the equilibrium bridge population $P_B^\infty$ of a
three-state system at a fixed bridge energy $\epsilon/\Delta=5$ for varying
temperature. As can be seen by comparing $P_B^\infty$ obtained from
imaginary-time MC simulations with a simple Boltzmann distribution,
i.e.~$1/[2\exp(\beta\epsilon)+1]$, deviations occur for $\hbar\beta\Delta >
0.3$ related to the strong influence of quantum delocalization (see
Fig.~\ref{fig_2}). This triggers the tendency of coherent oscillations in the
population dynamics which becomes especially obvious for an energetically
degenerated bridge (see Fig.~\ref{fig_3}). Damped oscillations are clearly
observable in the intermediate state for $\hbar\beta\Delta > 0.3$. These
results verify that (i) with the above choice of parameters electronic coherent
effects are crucial and (ii) that the ET dynamics can be captured by transfer
rates for $\hbar\beta\Delta < 0.3$. With respect to the bath, a typical
temperature $\hbar\beta\Delta = 0.1$ corresponds at the maximum $\omega_{\rm
max} = \omega_c$ of the spectral density to $\hbar\beta\omega_{\rm max} = 0.5$
so that a high temperature approximation does not apply.  The simulations have
been performed for system times long enough and sufficient stochastical
accuracy to allow for studying details of the dynamics even close to the
approach of equilibrium.

\begin{figure}
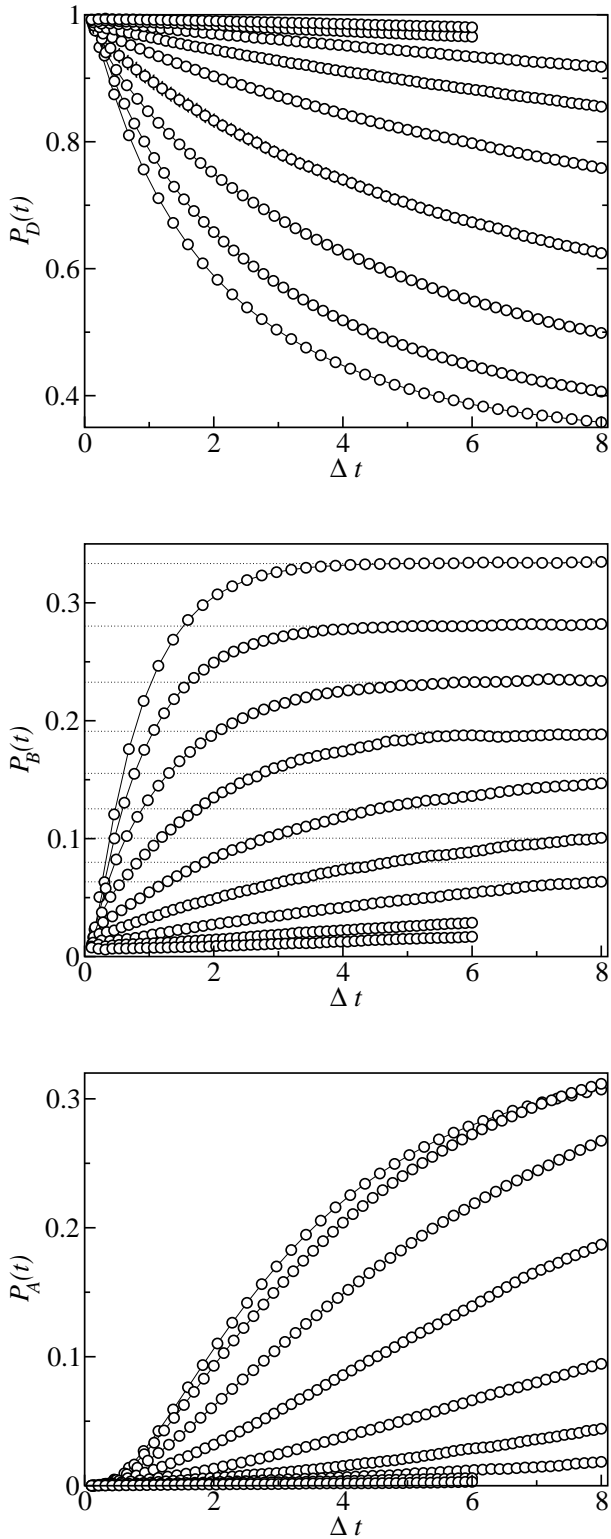

\epsfxsize=8cm
\epsffile{fig_4a.eps}\\[0.8cm]
\epsfxsize=8cm
\epsffile{fig_4b.eps}\\[0.8cm]
\epsfxsize=8cm
\epsffile{fig_4c.eps}
\caption[]{\label{fig_4} Donor, bridge, and acceptor populations $P_D(t)$,
 $P_B(t)$, and $P_A(t)$, respectively (from top to bottom), for $S = 1$,
 $\alpha = 1/4$, $\omega_c/\Delta = 5$, $\hbar\beta\Delta = 0.1$, and (top to
 bottom for bridge and acceptor, bottom to top for donor) $\epsilon/\Delta$ =
 0, 2.5, 5, 7.5, 10, 12.5, 15, 17.5, and 20. Error bars correspond to one
 standard deviation in each direction. Dotted lines denote the respective
 Boltzmann equilibrium populations of the bridge state. The solid lines are
 guides for the eye only.}
\end{figure}

\begin{figure}
\epsfxsize=8cm
\epsffile{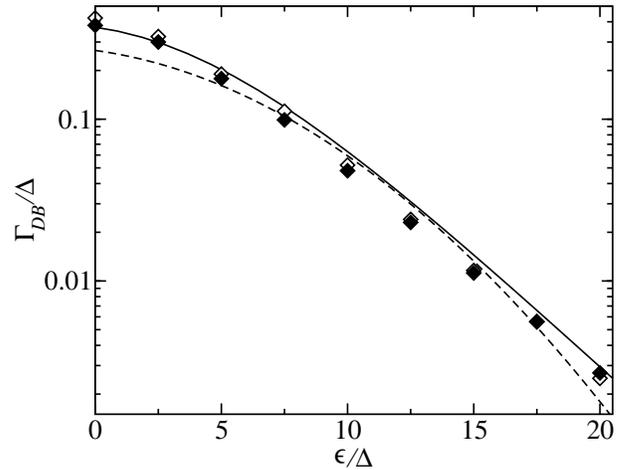}
\caption[]{\label{fig_5} Sequential transfer rates $\Gamma_{DB}$ obtained from
  the bridge state (open diamonds) and donor and acceptor state (closed
  diamonds) for $S = 1$, $\alpha = 1/4$, $\omega_c/\Delta = 5$, and
  $\hbar\beta\Delta = 0.1$ as a function of $\epsilon$. The solid line refers
  to the golden rule rate (\ref{grs}), the dashed one to the Marcus rate
  (\ref{marcus}).}
\end{figure}

Having fixed the proper range of parameters, we now turn to a detailed analysis
of the population dynamics. First, we consider a three-state system with
varying bridge energy $0\leq \epsilon/\Delta\leq 20$ and at a fixed temperature
$\hbar\beta\Delta = 0.1$ (Fig.~\ref{fig_4}). For bridges with
$\epsilon/\Delta\leq 7.5$ the intermediate state approaches thermal equilibrium
within the simulation period. Interestingly, a bridge energy $\epsilon/\Delta =
2.5$ leads to an increase of the acceptor population almost as fast as for the
degenerate case $\epsilon/\Delta=0$. This is due to the fact that for the
former bridge the equilibrium population $P_A^\infty$ is larger while the
bridge is still not high enough to considerably reduce the donor-bridge rate
$\Gamma_{DB}$. The population dynamics for this system can now be analyzed with
the full (\ref{convME}) and the sequential (\ref{bridgeME}) model to extract
the rate constants for times $t > t_{\rm trans}$. The rate constants were then
obtained by repeating this procedure for successively increasing $t_{\rm
trans}$ such that the corresponding rates eventually saturate at plateau
values. The latter are depicted in Fig.~\ref{fig_5} where also the nonadiabatic
classical Marcus (\ref{marcus}) and the golden rule (\ref{grs}) results are
shown. The numerical results basically follow the latter ones with minor
deviations that can be attributed to adiabatic effects (e.g.~recrossing
phenomena) not included in the golden rule rate expression. Sequential rates
are extracted for all three electronic states where those obtained from the
donor and acceptor coincide while rates extracted from the bridge exceed the
former by about 5-10\%. This deviation signals a shortcoming of
Eqs.~(\ref{bridgeME}) and (\ref{matrix}) in this parameter regime. In fact, the
reduction of Eq.~(\ref{exactME}) to Eq.~(\ref{convME}) is in a strict sense
only justified in the strongly damped regime. The impact of superexchange,
however, is very weak and its corresponding rates are smaller by two orders of
magnitude. Accordingly, the sequence of superexchange rates for various $t_{\rm
trans}$ obtained from the fitting procedure described above does not saturate
but rather strongly fluctuates such that a quantitative description is out of
range.

This behavior can be understood from the analytical rate expressions. Namely,
upon closer inspection of Eqs.~(\ref{super2}) and (\ref{marcus}) one derives
that in the {\sl classical regime} $\Gamma_{SCL, \rm GR}$ dominates against the
sequential transfer $\Gamma_{DB, \rm Marcus}$ whenever $\hbar\beta\epsilon\gg
1$ and $\epsilon\gg \Lambda_{\rm cl}$ such that
\begin{equation} \label{compa1}
\frac{\Gamma_{SCL, \rm GR}}{\Gamma_{DB, \rm Marcus}}\approx
\frac{\Delta^2}{\epsilon^2} \exp(\beta\hbar\epsilon^2/4\Lambda_{\rm
cl})\gg 1\; .
\end{equation}
This verifies the expected and well-known fact that for a classical bath the
superexchange mechanism prevails whenever thermal activation from D to B is
suppressed. For a {\sl quantum bath}, however, the situation is more
restrictive. From Eq.~(\ref{super1}) one estimates $\Gamma_{SQM, \rm GR}
\propto \Delta^4/(\omega_c \epsilon^2)$ for $\hbar\beta\omega_c > 1$ (and
$\alpha$ of order 1) such that for fixed temperature $\hbar\beta\Delta < 0.5$
superexchange transport prevails only for extremely high bridge energies,
i.e. $\epsilon/\Delta > 40$, where the transfer is essentially frozen on the
time scale accessible in MC simulations and in realistic experiments. With
electronic couplings of the order of 100$\,$cm$^{-1}$ \cite{friesner,nitzan1},
the above condition requires at $T \simeq 150\,$K bridge energies of the order
of $\epsilon \approx 5000$$\,$cm$^{-1}$ leading to typical transfer times of
the order of 30$\,\mu$s. Having in mind that our primary focus is to identify
regions where one of the transfer channels, sequential or superexchange, {\em
dominates}, superexchange turns out to be negligible in Fig.~\ref{fig_4}.

\begin{figure}
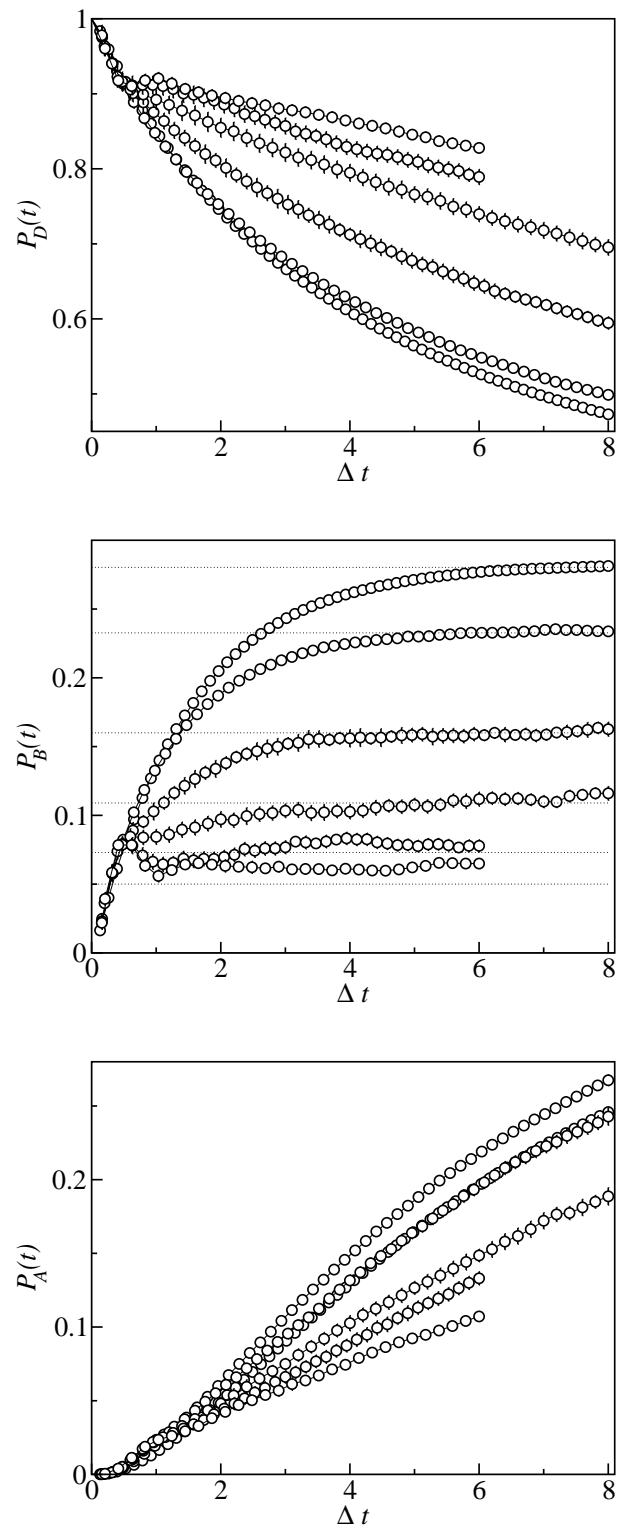

\epsfxsize=8cm
\epsffile{fig_6a.eps}\\[0.7cm]
\epsfxsize=8cm
\epsffile{fig_6b.eps}\\[0.7cm]
\epsfxsize=8cm
\epsffile{fig_6c.eps}
\caption[]{\label{fig_6} Donor, bridge, and acceptor populations $P_D(t)$,
 $P_B(t)$, and $P_A(t)$, respectively (from top to bottom), for $S = 1$,
 $\epsilon/\Delta = 5$, $\alpha = 1/4$, $\omega_c/\Delta = 5$, and (top to
 bottom for bridge and acceptor, bottom to top for donor) $\hbar\beta\Delta$ =
 0.05, 0.1, 0.2, 0.3, 0.4, and 0.5. Error bars correspond to one standard
 deviation in each direction. Dotted lines denote the respective equilibrium
 populations from Fig.~\ref{fig_2} of the bridge state. The solid lines are
 guides for the eye only.}
\end{figure}

\begin{figure}
\epsfxsize=8cm
\epsffile{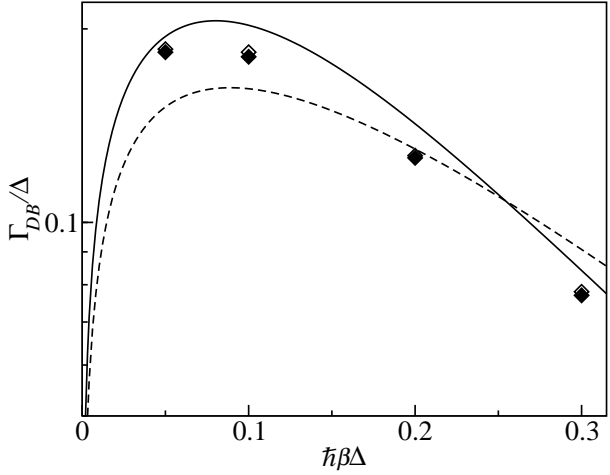}
\caption[]{\label{fig_7} Sequential transfer rates $\Gamma_{DB}$ obtained from
  the bridge state (open diamonds) and donor and acceptor state (closed
  diamonds) for $S = 1$, $\epsilon/\Delta = 5$, $\alpha = 1/4$, and
  $\omega_c/\Delta = 5$ as a function of inverse temperature. The solid line
  refers to the golden rule rate (\ref{grs}), the dashed one to the Marcus rate
  (\ref{marcus}).}
\end{figure}

At least as a minor effect we find superexchange in the ET dynamics shown in
Fig.~\ref{fig_4} for $\epsilon/\Delta \geq 5$. We thus performed simulations
for fixed bridge energy, but varying temperature $0.05 \leq \hbar\beta\Delta
\leq 0.5$, see Fig.~\ref{fig_6}. As already mentioned, for $\hbar\beta\Delta
> 0.3$ a rate description is questionable as the bridge population shows
signs of coherent oscillations. Sequential rates are extracted according to the
above scheme from donor and acceptor as well (cf.~Fig.~\ref{fig_7}). The
results are in qualitative agreement with the golden rule description with
deviations between donor and acceptor rates appearing in the low temperature
range where quantum coherences exist. All numerical rates are below the
nonadiabtic values, again due to adiabatic effects in the bath
dynamics. Superexchange cannot be found. This is understood from the above
discussion: For fixed $\epsilon$ the analytical superexchange rate exceeds the
golden rule one only in the temperature range $\hbar\beta\Delta \gg 2
\ln(\epsilon/\Delta)/(\epsilon/\Delta)$ corresponding to low temperatures
$\hbar\beta\Delta \gg 0.6$ where, as seen above, for weaker dissipation
($\alpha < 1/2$) a rate description is in a strict sense no longer feasible.
The situation changes when donor and acceptor are biased with an energy gap
$\epsilon_{DA} = \epsilon_D-\epsilon_A > 0$. Then, based on the generalization
of Eq.~(\ref{super1}) for $\epsilon_{DA} \ne 0$ \cite{nica} one can show for a
high lying intermediate state that superexchange exceeds sequential transport
already at higher temperatures $\hbar\beta\Delta\gg 2 {\rm
ln}(\epsilon/\Delta)/[(\epsilon_{DA}+\epsilon)/\Delta]$. In fact, for biased ET
systems indications of superexchange have already been seen in previous MC
simulations for three-state systems \cite{optimized filter}.

\begin{figure}
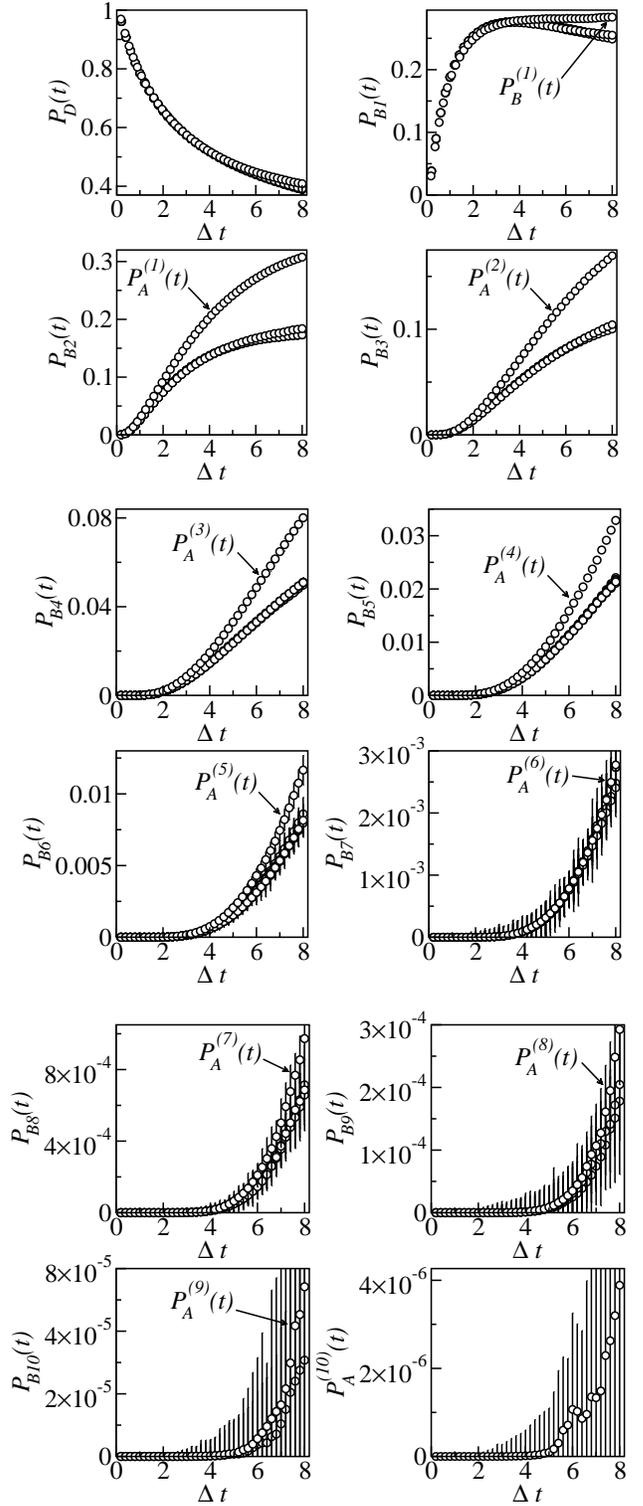

\epsfxsize=7.6cm
\hspace*{3mm}\epsffile{fig_8a.eps}\\[0.3cm]
\epsfxsize=8cm
\epsffile{fig_8b.eps}\\[0.3cm]
\epsfxsize=8.2cm
\hspace*{-1mm}\epsffile{fig_8c.eps}
\caption[]{\label{fig_8} Electronic populations for $\epsilon/\Delta = 2.5$,
  $\alpha = 1/4$, $\omega_c/\Delta = 5$, $\hbar\beta\Delta = 0.1$, and $1 \le S
  \le 3^{ 1\!}/_2$ $(b=1, \ldots, 10)$. $P_A^{(b)}$ denotes the acceptor
  population for a system with $b$ bridge states.}
\end{figure}

Let us now turn to longer bridges with $b = d-2 > 1$. The stability of the MC
procedure allows to extract rate constants for systems up to $b = 10$. In
Fig.~\ref{fig_8} results for $b = 1$ up to $b = 10$ bridge states are
displayed. Apparently, the donor dynamics basically saturates for $b > 2$ on
the time scale of the simulations meaning that the donor decay depends on the
bridge length only at very large times and then only slightly. This amounts to
the fact that the equilibrium populations decrease only algebraically with
$b$. In contrast, the time evolution of the acceptor and its closest bridge
states are strongly affected by the varying bridge length. The numerical data
can be very precisely reproduced by a sequential transfer model according to
Eq.~(\ref{bridgeME}) where due to symmetry and detailed balance only two fit
parameters enter, namely the forward rate from the donor $\Gamma_{DB}$ and the
bridge rate $\Gamma_B$. It turns out that by fixing these parameters at
$\Gamma_{DB} = 0.284$ and $\Gamma_B = 0.348$ {\em all} populations independent
of the bridge length can be described, thus verifying that the transport is
more or less completely sequential. The nonadiabatic golden rule formula
(\ref{grs}) predicts a little bit higher values $\Gamma_{DB,GR} = 0.297$ and
$\Gamma_{B,GR} = 0.37$ in agreement with our findings for the three-state
system.
Our main focus lies on the behavior of the smallest eigenvalue of the rate
matrix $A$, i.e.~the relaxation rate $\Gamma_R$, as the bridge length $b$
increases. Since an analytical diagonalization of $A$ for arbitrary $b$ is not
possible, we invoke an effective DBA model to capture the key features and
compare with the numerical results. Accordingly, we follow Ref.~\cite{pollak}
and consider in addition to the donor and the acceptor population only the
bridge populations $P_{BD}$ and $P_{BA}$ coming from the donor and acceptor,
respectively. The dynamics of these populations ${\bf \cal P}=(P_D, P_{BD},
P_{BA},P_A)$ is then determined by $\dot{{\bf \cal P}}={\cal A}\, {\bf \cal P}$
with the reduced rate matrix
\begin{equation} \label{simrate}
{\cal A} = \left(
\begin{array}{cccc}
-\Gamma_{DB} & \Gamma_r & \Gamma_d & 0\\
\Gamma_{DB} & -\Gamma_r-\Gamma_d & 0 &0 \\
0 &0 &-\Gamma_r-\Gamma_d & \Gamma_{DB}\\
0 & \Gamma_d &\Gamma_r & -\Gamma_{DB}
\end{array}
\right)\; .
\end{equation}
Here, $\Gamma_r$ denotes the recrossing rate from the bridge back to
donor/acceptor and $\Gamma_d$ the rate for the diffusive motion from one end to
the other end of the bridge. The smallest eigenvalue can now be found
analytically and reads
\begin{equation} \label{brelax}
\Gamma_R = \frac{\Gamma_{DB}}{2}\
\left[1+\mu_r+\mu_d-\sqrt{(1+\mu_r+\mu_d)^2-8 \mu_d}\right]
\end{equation}
where for convenience we introduced $\mu_{r/d} = \Gamma_{r/d}/\Gamma_{DB}$.  In
Ref.~\cite{pollak} it was found that $\mu_r/\mu_d = b$. Since $\Gamma_{DB}$ is
independent of the bridge length, the ratio $\mu_d$ can be inferred from the
mean first passage time of a particle diffusing across a one-dimensional wire
\cite{landauer}. This scales like $b^2$ so that $\mu_d=1/(\nu\, b^2)$ where
$\nu$ is a constant. Based on these results two scenarios for $\Gamma_R(b)$ can
be distinguished. For short and moderate long bridges with $b$ sufficiently
smaller than $1/\nu$ ($\mu_d > 1$), Eq.~(\ref{simrate}) gives rise to
$\Gamma(b) \approx \Gamma_{DB} \ 2/(1+b)$, while for longer bridges $b \gg
1/\nu$ ($\mu_d \ll 1$) the approximation $\Gamma_R(b) \approx \Gamma_{DB}\
2/(\nu\, b^2)$ applies. In the first case the bottleneck of the transfer
process is the donor-bridge activation encoded in $\Gamma_{DB}$, in the latter
one the rate is limited by the diffusive motion along the entire bridge
described by $\Gamma_d$. An effective formula comprising these limiting cases
is given by
\begin{equation} \label{long}
\Gamma_R(b)\approx \Gamma_{DB}\, \frac{2}{1+b+\nu \, b^2}\; .
\end{equation}

\begin{figure}
\epsfxsize=8cm
\epsffile{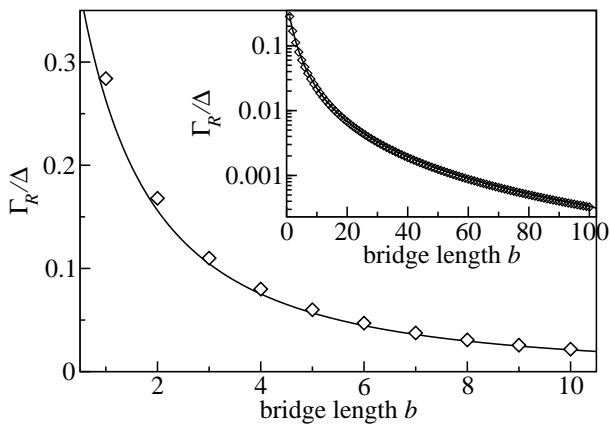}
\caption[]{\label{fig_9} Relaxation rate $\Gamma_R$ for $\epsilon/\Delta =
  2.5$, $\alpha = 1/4$, $\omega_c/\Delta = 5$, $\hbar\beta\Delta = 0.1$, and $1
  \le S \le 3^{ 1\!}/_2$ ($1 \le b \le 10$) (open diamonds). The solid line
  resembles Eq.~(\ref{long}). The rates in the inset correspond to the golden
  rule rate (\ref{grs}).}
\end{figure}

Indeed, the numerical data shown in Fig.~\ref{fig_9} are quantitatively
captured by the result (\ref{long}) with $\nu = 0.159$. Due to this
excellent agreement between sequential transfer model and the PIMC data on the
one hand and between the corresponding relaxation rates and the expression
(\ref{simrate}) on the other hand, we extracted the relaxation rate from the
full rate matrix $A$ for bridges with up to $b = 100$ and confirmed the
validity of Eq.~(\ref{simrate}).

A functional behavior of the form $\Gamma_R(b) \propto 1/b$ has been noted
empirically already in Ref.~\cite{nitzan1}, derived theoretically in
Ref.~\cite{pollak} (in a slightly different model though), and observed
experimentally in a variety of molecular systems
\cite{joachim,ratner_nature,nitzan2}. In lead-molecule-lead junctions it gives
rise to Ohm's law in the current voltage characteristics. Our above findings
indicate that, at least in symmetric DBA systems, this behavior is
characteristic for situations where $\Gamma_{DB}$ is sufficiently smaller than
the bridge diffusion rate. However, a changeover to $\Gamma(b)\propto 1/b^2$
should be observable for very long bridges.

\section{Conclusions}
\label{Sec: Conclusions}

Electron transfer processes across molecular chains have gained much interest
in recent years, particularly in the context of molecular
electronics. Motivated by previous studies, in this paper the electronic
dynamics across symmetric donor-bridge-acceptor systems has been analyzed based
on a numerically exact treatment of the corresponding dissipative quantum
system within a real-time MC scheme.

Upon improving existing filter techniques we have been able to achieve
accurate data for the population dynamics even for times where equilibration
sets in or has already been established, and for an increasing number of
electronic bridge states. This opens the door to consider a variety of
electronic site topologies as well as the impact of external time dependent
fields. Eventually, by a proper reduction of the Hilbert space also the reduced
dynamics of continuous degrees of freedom with a discrete energy spectrum of
the bare system are now accessible. Corresponding work in this direction is in
progress.

Physically, the main issue has been the role of superexchange and sequential
transport in symmetric DBA compounds. Here, results based on Redfield theory
and simplified system-bath couplings indicated different regimes for the length
dependence of net transfer across the bridge comprising an exponential and an
algebraic fall-off, respectively.
The numerical MC data have been analyzed by exploiting the fact that (i) the ET
dynamics can in the range of parameters considered be mapped onto Master
equations with time independent transition rates and (ii) the path integral
formulation reproduces the known analytical findings for these rates in certain
limits. For the DBA compound the parameters were chosen to guarantee strong
quantum effects, i.e.~moderate dissipation, sufficiently low temperatures, and
a broad distribution of bath modes.

The conclusion of the numerical results together with the analytical findings
is that in a symmetric DBA system (degenerate donor and acceptor energy) even
for weaker dissipation superexchange can be expected to dominate only for
classical or nearly classical bath. For quantum mechanical baths as they have
been studied here, it does not play a significant role for ET processes that
can be captured by time independent transport rates. Particularly, in contrast
to earlier studies we found no dominant role of the superexchange mechanism
neither for relatively high lying bridges, nor for lower temperatures or more
bridge sites.
For quantum baths it requires extremely high lying bridge states with
essentially frozen dynamics or temperatures that are so low that coherences
give rise to oscillating population dynamics associated with a breakdown of a
conventional rate description. Experimentally, a clear observation of a
changeover from tunneling to hopping dominated transfer in molecular chains has
not been found yet, a fact, that may to some extent be attributed to the
scenario we have revealed. The results reported in Ref.~\cite{ratner_nature}
show some indications but not a direct proof though, since the energetic
landscape of the molecular structures changes with their length.
By varying the number of bridge sites an algebraic decrease of the relaxation
rate could be confirmed for shorter as well as longer bridges.  Based on a
reduced DBA model we find a changeover from a range where the donor-bridge
activation limits the transfer to a domain where the nondirectional diffusive
motion along the wire is decisive. Previous simulations on biased three-state
systems suggest in accordance with the analytical results that an energy gap
between donor and acceptor (in addition to the energy gap between the bridge
and donor/acceptor) stimulates the occurrence of superexchange.

\section*{Acknowledgment}

Financial support was provided by the DFG (Bonn) through AN 336-1 and a
Heisenberg fellowship for JA.

\appendix
\section{Computational details}
\label{App: Computational details}

For the sake of computational savings, we calculate
\begin{eqnarray} \label{pop-disc}
P_{s_f,s_i}(t_k) &=& {1\over2} \sum_{\{s_j\}} \left(\delta_{s_k,s_f} +
\delta_{s'_k,s_f}\right) \, \delta_{s_1,s_i} \rho[\{s_j\}]\;, \nonumber\\
&& k=1,\ldots, q+1
\end{eqnarray}
rather than Eq.~(\ref{pop-discretized}), such that one single MC
trajectory can be used to compute $P_{s_f,s_i}$ for all times $t_k =
(k-1)\tau$, where the simultaneous measurement on the forward and
backward path enhances the statistics. Equation~(\ref{population}) then
becomes
\begin{eqnarray}
P_{s_f, s_i}(t) &=& {1\over2}\sum_{\{\xi_j\}, \eta_{q+1}}
 \exp\!\left[-\sum_{j=2}^q \xi_j \Bigg(i\hat{X}^{(s_i)}_j \right. \nonumber\\
&&\left.\left.{} +\sum_{k \ge j} \xi_k \Lambda_{k-j} \right)\right]\hat{K}^{(1)}(\xi_2, \dots, \xi_q) \cdots \nonumber\\
&& \times
 \bar{K}^{(j;s_f)}(\xi_j, \dots, \xi_q) \cdots \hat{k}^{(q)}(\xi_q, \eta_{q+1})\;, \nonumber\\
\end{eqnarray}
where $\bar{K}^{(j;s_f)}(\xi_k, \dots, \xi_q)$ is defined by
\begin{eqnarray}
\bar{K}^{(j;s_f)}_{\eta_j,\eta_{j+1}}(\xi_j, \dots, \xi_q) &=&
{1\over2}\left(\delta_{\eta_k+\xi_k,2s_f} +
\delta_{\eta_k-\xi_k,2s_f}\right)
\nonumber\\
&&\times \hat{K}^{(j)}_{\eta_j,\eta_{j+1}}(\xi_j,\dots,\xi_q)
\end{eqnarray}
and $\hat{K}^{(1)}$ again is a row vector due to $\eta_1 \equiv 2s_i$,
while $\hat{k}^{(q)}(\xi_q, \eta_{q+1})$ denotes the $\eta_{q+1}$'s
column of $\hat{K}^{(q)}(\xi_q)$.

Another significant reduction of the computational costs of the MC calculations
can be obtained by optimizing the MC weight. While
\begin{eqnarray} \label{MC-weight 1}
\lefteqn{w_{\rm MC}(\{\xi_j\}, \eta_{q+1}) \equiv
 \exp\!\Bigg(-\sum_{k \ge j = 2}^q \xi_k \Lambda_{k-j}
 \xi_j \Bigg)} \nonumber\\
&& \times \left|\hat{K}^{(1)}(\xi_2,\dots,\xi_q) \cdots
 \hat{k}^{(q)}(\xi_q, \eta_{q+1})\right|
\end{eqnarray}
generally represents a natural choice for the MC weight, for systems with
sufficiently weak damping an attractive alternative is given by
\begin{equation} \label{MC-weight 2}
\tilde{w}_{\rm MC}(\{\xi_j\}, \eta_{q+1}) \equiv
 \left|\tilde{K}^{(1)}(0,\xi_2) \cdots \tilde{k}^{(q)}(\xi_q)\right|
\end{equation}
with the real-valued matrices
\begin{equation}
\tilde{K}_{\eta,\eta'}(\xi,\xi') = |\langle (\eta+\xi)/2|\exp(-i\tau
H_{d\rm LS}/\hbar)|(\eta'+\xi')/2\rangle| \;,
\end{equation}
which essentially neglect all dissipative effects. Although MC runs utilizing
$\tilde{w}_{\rm MC}$ exhibit somewhat poorer statistics than those using
$w_{\rm MC}$, the switching from complex to real-valued matrices which
furthermore, unlike the $\hat{K}^{(j)}$, can be calculated and stored before
executing the MC moves results in a significant speedup of the simulations. In
fact, for the parameters investigated here, using Eq.~(\ref{MC-weight 2}) over
(\ref{MC-weight 1}) could reduce the computational costs by a factor of roughly
four. Note that the restriction of the use of $\tilde{w}_{\rm MC}$ to the
weakly damped regime does not really impose a severe limitation since for
sufficiently strong dissipation the overall sign problem will be weak enough to
be tackled by any conventional PIMC scheme.

\section{Population dynamics for ${\mathbf S=1}$}
\label{App: Expressions for the populations}

For $S=1$, explicit and transparent expressions for the electronic populations
can be gained from Eq.~(\ref{bridgeME}) and (\ref{matrix}). Allowing some
transient motion for times $t < t_{\rm trans}$, one easily obtains
\begin{eqnarray} \label{explicit population}
\lefteqn{P_D(t_{\rm trans}+t) =} \nonumber\\
&& {1\over2} \Bigg\{ 1
  + [P_D(t_{\rm trans}) - P_A(t_{\rm trans})]e^{-(\Gamma_{DB} + \Gamma_S)t} \nonumber\\
&&\quad{} - [P^\infty_B - P_B(t_{\rm trans})]\left[1 -
  \exp\!\left({}-{\Gamma_{DB} \over P_B^\infty}t\right)\right] \nonumber\\
&&\quad{} - P_B(t_{\rm trans})\Bigg\} \;,\nonumber\\
\lefteqn{P_B(t_{\rm trans}+t) =} \nonumber\\
&& [P_B^\infty - P_B(t_{\rm trans})]\left[1 - \exp\!\left({}-{\Gamma_{DB} \over
 P_B^\infty}t\right)\right]\nonumber\\
&&\quad{} + P_B(t_{\rm trans}) \;, \nonumber\\
\lefteqn{P_A(t_{\rm trans}+t) =} \nonumber\\
&& {1\over2} \Bigg\{ 1
  + [P_A(t_{\rm trans}) - P_D(t_{\rm trans})]e^{-(\Gamma_{DB} + \Gamma_S)t} \nonumber\\
&&\quad{} - [P^\infty_B - P_B(t_{\rm trans})]\left[1 -
\exp\!\left({}-{\Gamma_{DB}
  \over P_B^\infty}t\right)\right] \nonumber\\
&&\quad{} - P_B(t_{\rm trans})\Bigg\} \;,
\end{eqnarray}



\begin{thebibliography}{99}
%
\bibitem{jortner} J. Jortner and M. Bixon, eds., Adv. Chem. Phys. {\bf
106}, 107 (1999).
%
\bibitem{marcus56} R.A. Marcus, J. Chem. Phys. {\bf 24}, 966 (1956).
%
\bibitem{marcus85}
R.A. Marcus and N. Sutin, Biochim. Biophys. Acta {\bf 811}, 265 (1985).
%
\bibitem{friesner} A.K. Felts, W.T. Pollard, and R.A. Friesner, J. Phys.
Chem. {\bf 99}, 2929 (1995).
%
\bibitem{mukamel} A. Okada, V. Chernyak, and S. Mukamel, J. Phys. Chem. A
{\bf 102}, 1241 (1997).
%
\bibitem{nitzan0} W.B. Davis, M.R. Wasielewski, M. Ratner, V. Mujica, and
A. Nitzan, J. Phys. Chem. {\bf 101}, 6158(1997).
%
\bibitem{ratner_nature} W.B. Davis, W.A. Svec, M.A. Ratner, and M.R.
Wasielewski, Nature {\bf 396}, 60 (1998).
%
\bibitem{nitzan1} D. Segal, A.Nitzan, W.B. Davis, M.R. Wasielewski, and
M.A. Ratner, J. Phys. Chem. {\bf 104}, 3817 (2000).
%
\bibitem{pollak} E. Hershkovitz and E. Pollak, Ann. Phys. (Leipzig) {\bf
9}, 764 (2000).
%
\bibitem{dekker} C. Dekker and M.A. Ratner, Phys. World {\bf 14}, 29
(2001).
%
\bibitem{joachim} M. Magoga and C. Joachim, Phys. Rev. B {\bf 56}, 4722
(1997).
%
\bibitem{ratner} J. Jortner and M.A. Ratner (eds.), {\em Molecular
Electronics}, (Blackwell Sci., Oxford, 1997).
%
\bibitem{saclay} C. Kergueris, J.-P. Bourgoin, S. Palacin, D. Esteve, C.
Urbina, M. Magoga, and C. Joachim, Phys. Rev. B {\bf 59}, 12505 (1999).
%
\bibitem{nitzan2} A. Nitzan, Ann. Rev. Phys. Chem. {\bf 52}, 681 (2001).
%
\bibitem{hanggi} P. H{\"a}nggi and M.A. Ratner (eds.), Chem. Phys. {\bf
281}, (2002).
%
\bibitem{chandler}
D. Chandler, in {\sl Liquids, Freezing and the Glass Transition}, ed.~by
D. Levesque, J.P. Hansen. and J. Zinn-Justin (Elsevier Science, North
Holland, 1991), Les Houches 51, Part 1.
%
\bibitem{song}
X. Song and A.A. Stuchebrukhov, J. Chem. Phys. {\bf 99}, 969 (1993).
%
\bibitem{weiss}
U. Weiss, {\sl Quantum Dissipative Systems}, Series in Modern Condensed
Matter Physics, Vol. 2 (World Scientific, Singapore, 1998).
%
\bibitem{feynman}
R.P. Feynman and F.L. Vernon, Ann. Phys. (N.Y.) {\bf 24}, 118 (1963).
%
\bibitem{leggett} A.O. Caldeira and A.J. Leggett, Ann. Phys. (NY) {\bf
149}, 374 (1983); {\em ibid.} {\bf 153}, 445(E) (1983).
%
\bibitem{makri} N. Makri and D.E. Makarov, J. Chem. Phys. {\bf 102}, 4600
(1995).
%
\bibitem{sign-problem}
 {\em Quantum Monte Carlo
Methods in Condensed Matter Physics}, ed.~by M. Suzuki (World Scientific,
Singapore, 1993), and references therein.
%
\bibitem{lucke1} A. Lucke, C.H. Mak, R. Egger, J. Ankerhold, J.T.
Stockburger, and H. Grabert, J. Chem. Phys. {\bf 118}, 291 (1997).
%
\bibitem{lucke2} A. Lucke and J. Ankerhold, J. Chem. Phys. {\bf 115}, 4696
(2001).
%
\bibitem{mlb03}
L. M{\"u}hlbacher and R. Egger, J. Chem. Phys. {\bf 118}, 179 (2003).
%
\bibitem{optimized filter}
R. Egger and C.H. Mak, Phys. Rev. B {\bf 50}, 15210 (1994); R. Egger and
C.H. Mak, J. Phys. Chem. {\bf 98}, 9903 (1994).
%
%
\bibitem{entropy}
S. Chakravarty and J. Rudnick, Phys. Rev. Lett. {\bf 75}, 501 (1995).
%
\bibitem{filter1}
V.S. Filinov, Nucl. Phys. B {\bf 271}, 717 (1986).
%
\bibitem{filter2}
See, for example, J.D. Doll and D.L. Freeman, in {\sl Lasers, Molecules,
and
 Methods}, ed.~by J.O. Hirschfelder, R.E. Wyatt, and R.D. Coalson,
Adv. Chem. Phys. Vol. LXXIII (Wiley, New York, 1989).
%
\bibitem{makegger}
C.H. Mak and R. Egger, Adv. Chem. Phys. {\bf 93},39 (1996).
%
\bibitem{mlb98}
C.H. Mak, R. Egger, and H. Weber-Gottschick, Phys. Rev. Lett. {\bf 81},
 4533 (1998); R. Egger, L. M{\"u}hlbacher, and C.H. Mak, Phys. Rev. E {\bf 61},
5961 (2000).
%
\bibitem{GR2}
R. Egger, C.H. Mak, and U. Weiss, J. Chem. Phys. {\bf 100}, 2651 (1994).
%
\bibitem{nica} R. Egger, C.H. Mak, and U. Weiss, Phys. Rev. E {\bf 50},
(655)(R) (1994).
%
\bibitem{hermann_talkner}
H. Grabert, U. Weiss, and P. Talkner, Z. Phys. B {\bf 55}, 87 (1984).
%
\bibitem{lehle} J. Ankerhold and H. Lehle, J. Chem. Phys. {\bf 120}, 1436
(2004).
%
\bibitem{weber} J. Reichert, R. Ochs, H.B. Weber, M. Mayor and, H.v. L\"ohneysen,
Appl. Phys. Lett. {\bf 82}, 4137 (2003).
%
\bibitem{landauer} R. Landauer and M. B{\"u}ttiker, Phys. Rev. B {\bf 36},
6255 (1987).
\end{thebibliography}
\end{document}